\documentclass[english,noshowpacs, amssymb,twocolumn,aps,nofootinbib]{revtex4}
\usepackage{graphicx}

\begin{document}

\title{Einstein-Podolsky-Rosen Paradox with Position-Momentum\\ Entangled Macroscopic Twin Beams}

\author{Ashok Kumar$^{1,2}$, Gaurav Nirala$^{1,3}$, and  Alberto M. Marino$^{1,3}$}
\affiliation{$^1$Homer L. Dodge Department of Physics and Astronomy, The University of Oklahoma, Norman, Oklahoma 73019, USA}
\affiliation{$^2$ Department of Physics, Indian Institute of Space Science and Technology, Thiruvananthapuram, Kerala, 695547, India}
\affiliation{$^3$Center for Quantum Research and Technology, The University of Oklahoma, Norman, Oklahoma 73019, USA}

\begin{abstract}
Spatial entanglement is at the heart of quantum enhanced imaging applications and high-dimensional quantum information protocols. In particular, for imaging and sensing applications, quantum states with a macroscopic number of photons are needed  to provide a real advantage over the classical state-of-the-art. We demonstrate the Einstein-Podolsky-Rosen (EPR) paradox in its original position and momentum form with bright twin beams of light by showing the presence of EPR spatial (position-momentum) entanglement. An electron-multiplying charge-coupled-device camera is used to record images of the bright twin beams in the near and far field regimes to achieve an apparent violation of the uncertainty principle by more than an order of magnitude. We further show that the presence of quantum correlations in the spatial and temporal degrees of freedom leads to spatial squeezing between the spatial fluctuations of the bright twin beams in both the near and far fields. This provides another verification of the spatial entanglement and points to the presence of hyperentanglement in the bright twin beams.
\end{abstract}

\maketitle

Over the last several decades, quantum entanglement has been studied extensively and is now considered to be an indispensable resource for the emerging field of quantum technologies~\cite{ent,Nielsen,Leuchs,Bouwmeester,Karlsson,Huelga,Kimble,Giovannetti,Marino,Marino2,LIGO,Dowran}.  Spatial entanglement, in particular, has attracted significant attention for applications in quantum information science as it exists in an infinite dimensional Hilbert space \cite{Howell, Padgett, Lantz, Lantz2, Lee,Kumar1, Fleischer,Wasilewski,Boyer}. An increased dimensionality can enable, for example, an exponential speed-up for quantum computation, enhanced quantum channel capacities, and security enhancements of quantum communication protocols~\cite{Bechmann, Wang, Huber,Mohammad,Krenn,Martin}.
Furthermore, spatial quantum correlations can extend quantum-based enhancements from the time domain to the spatial one to enable quantum imaging and quantum metrology with enhanced sensitivity and resolution \cite{Serg1, RMP, Treps1,Kolobov,Brida, Walborn, Genovese,Serg3}.

Spatial quantum correlations were central to the original Einstein-Podolsky-Rosen (EPR) paper of 1935~\cite{EPR} that questioned the completeness of quantum mechanics. The EPR paper considered a \textit{gedanken} experiment involving a pair of entangled particles with a space-like separation. The presence of perfect correlations between their positions and momenta led to an apparent violation of Heisenberg's uncertainty principle. Such an apparent violation is now known as the EPR paradox and it arises from imposing local realism for which two distant particles are treated as two independent systems~\cite{Bell, CHSS, Aspect, teleport, Banaszek}.

Experiments similar to the EPR thought experiment have now been performed with correlated photon pairs produced with parametric down conversion~\cite{Howell, Padgett, Lantz, Lantz2}. Initial experiments were performed in the time domain through the use of slits to select different spatial regions and temporal coincidence measurements with avalanche photodiodes~\cite{Howell}. More direct measurements of the spatial (position-momentum) quantum correlations were later performed with an electron-multiplying charge-coupled-device (EMCCD) camera~\cite{Lantz,Lantz2,Padgett}. EPR entanglement was  demonstrated by a verification of the EPR paradox through the violation of an inequality equivalent to Heisenberg's uncertainty principle~\cite{Reid, Reid2}.

While previous experiments have provided significant insight into the nature of  spatial quantum correlations, they have limited applicability for applications such as quantum sensing and imaging. This is due to the long integration times and/or large number of images required to observe quantum effects. The ability to generate and measure macroscopic quantum states that exhibit EPR spatial entanglement would make it possible to overcome such limitations.  In addition, it would boost the sensitivity of a given measurement due to the scaling of the signal-to-noise ratio with the number of photons. In practice,  to surpass the classical-state-of-the-art and make a difference for real-life applications, quantum states need to have a power close to the threshold power limit of the system to be enhanced. Once this limit is reached, further enhancements can only be obtained with quantum resources.  To this extent, in-vivo imaging of dynamic biological samples, which have a low damage threshold, provides an ideal application for macroscopic quantum states with spatial quantum correlations, as imaging over extended periods of time is not an option.

Here we demonstrate the EPR paradox in its original position-momentum form with macroscopic entangled beams of light, or bright twin beams, through measurements with an EMCCD. The photon flux of the bright twin beams is $\sim$10$^{14}$ photons per second per beam and is limited by the saturation of the EMCCD.  We show that the twin beams are EPR position-momentum entanglement through a violation of the EPR criterion by more than an order of magnitude and that a statistically significant violation is possible with $<10$ images. Moreover, we show that an interplay between quantum correlations in different degrees of freedom, spatial and temporal, leads to sub-shot noise spatial noise statistics, i.e. spatial squeezing.  The presence of spatial squeezing in both the near and far fields provides an additional verification of spatial entanglement in the twin beams.

To demonstrate the EPR paradox, the measured relative uncertainties in position and momentum between the twin beams must show an apparent violation of Heisenberg's uncertainty, which can be quantified through a violation of the EPR criterion
\begin{equation}
 \Delta^{2}r\Delta^{2}p \geq \hbar^{2}/4,
\label{Heisenberg}
\end{equation}
where $\Delta^{2}r=(\Delta r)^{2}$ and $\Delta^{2}p=(\Delta p)^{2}$ with $\Delta r$ and $\Delta p$ representing the position and momentum uncertainties, respectively, of one of the twin beams conditional on measurements of the corresponding variable on the other beam. Direct characterization of $\Delta r$ and $\Delta p$ is possible via spatial cross-correlation measurements between captured images of the twin beams in the near and far field regimes, respectively.

\begin{figure}[hbt]
\centering
\includegraphics{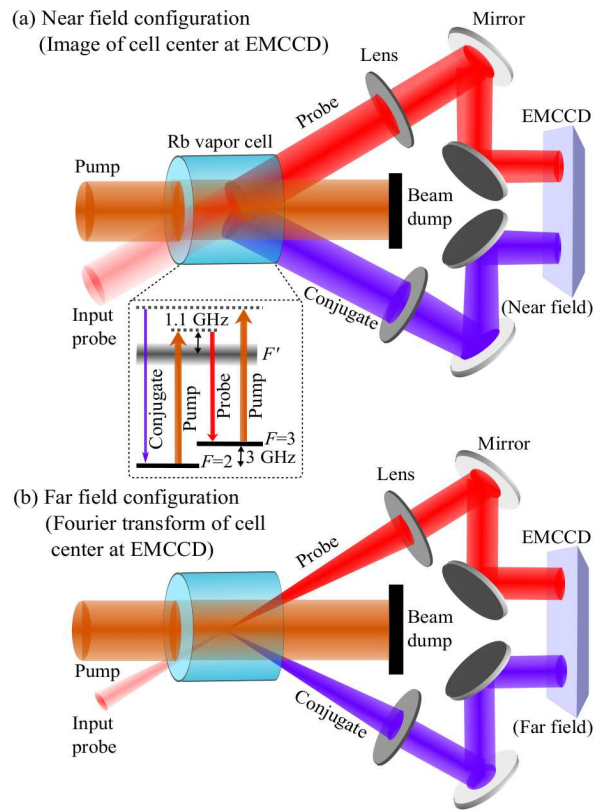}
\caption{Experimental setup to measure (a) position and (b) momentum correlations. For the position correlation measurements (near field) a single lens images the cell center onto the EMCCD, while for the momentum correlation measurements (far field) a single lens in an $f$-to-$f$ configuration generates the fourier transform of cell center at the EMCCD. The inset shows the double-$\Lambda$ energy level configuration in the D1 line of $^{85}$Rb used for the FWM.}
\label{Fig1}
\end{figure}

To measure the near and far field properties of the twin beams, we use the experimental setups shown in Figs.~\ref{Fig1}(a) and \ref{Fig1}(b), respectively. In both setups the positions of the source (Rb vapor cell) and EMCCD are kept fixed while different optical systems are used for each configurations. For the near field we image the cell center to the EMCCD with a 400~mm lenses in a configuration with a demagnification of 0.65. For the far field a 500~mm lens in an $f$-to-$f$ configuration generates the Fourier transform of the cell center on the EMCCD. The $f$-to-$f$ optical system maps the transverse momenta of the field at the cell center to transverse position on the EMCCD, such that a photon with transverse momentum $\hbar k_\perp$ is mapped to transverse position $x=f k_\perp/k$ in the far field~\cite{Howell}. For both configurations independent optical systems are used for the probe and conjugate beams.

We generate bright twin beams with a four-wave mixing (FWM) process in a double-$\Lambda$ configuration in the D1 line of $^{85}$Rb, see inset in Fig.~\ref{Fig1}. We use a Ti:Sapphire laser at 795~nm to generate the strong pump beam required for the FWM. An acousto-optic modulator is used to red-shift a portion of the laser by $\sim$3~GHz to generate the input  probe. Pump and probe are then made to intersect at an angle of 0.4 degrees inside a 12~mm long hot $^{85}$Rb vapor cell (temperature of 106$^\circ$C). In this configuration, two pump photons are absorbed and two new quantum correlated twin photons called probe and conjugate are generated. When seeded, the FWM amplifies the input probe beam and generates a bright conjugate beam to produce bright twin beams. By changing the number of photons in the input probe it is thus possible to obtain a controllable number of quantum correlated photons. This makes it possible to overcome some of the problems that limit the squeezing and number of photons in bright squeezed vacuum states \cite{Chekhova,Chekhova2}.

The diameter of the pump beam at the cell center is kept fixed at 4.4~mm; however, the input probe beam diameter at the cell center is set to 2.0~mm and 0.4~mm for the near and far field configurations, respectively. Such a change in size is needed to detect a large number of coherence areas (spatial modes) with the EMCCD, which is necessary to observe the spatial correlations~\cite{Marcelo}. The input probe beam acts as a seed to stimulate the generation of a macroscopic number of photons in the spatial modes it overlaps with.  As such, for the near (far) field configuration it is necessary for the probe diameter to be as large (small) as possible at the cell to excite as many spatial regions ($k$-vectors) as possible.

The technical details of the data acquisition are discussed in detail in our earlier work \cite{Kumar2}, here we only provide the relevant technical parameters. To acquire images of the bright probe and conjugate beams with the EMCCD, we pulse the input probe and pump beams with a duration of 1~$\mu$s and 10~$\mu$s, respectively. The probe pulse is delayed by 6~$\mu$s with respect to the pump pulse to avoid transients effects in the FWM. Acquisition of the probe and conjugate images with the EMCCD camera is synchronized with the pump-probe pulse timing sequence. We acquire 200 images, each with multiple frames, of the twin beams in both the near and far field configurations to observe the EPR paradox in position-momentum. We also acquire background images without an input probe beam after every probe-conjugate image acquisition to subtract the background noise due to electronic noise and scattered pump photons as done in Ref.~\cite{Kumar2}.

For bright optical fields, the quantum properties are present in their fluctuations and not in their mean values.  To obtain the spatial intensity fluctuations, we subtract two probe images and two conjugate images captured in consecutive frames (170$\times$512~pixels) with a time delay of 60~$\mu$s between them. Such a differential analysis technique extracts the spatial fluctuations of the images and also cancels the low frequency technical noise present in the twin beams. Given that the images in consecutive frames are taken in a time scale longer than the inverse of the bandwidth of the process, there are no quantum correlations between them.

\begin{figure}
\centering
\includegraphics[width=\linewidth]{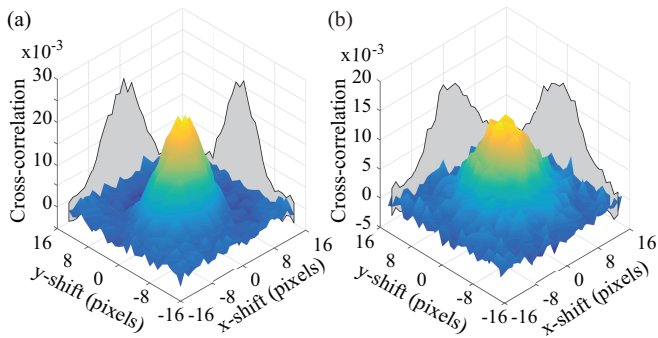}
\caption{Spatial cross-correlations between the intensity fluctuations of the probe and conjugate measured in the (a) near and (b) far field configurations.}
\label{Fig2}
\end{figure}

To calculate the spatial cross-correlations, we crop a region of 120$\times$120~pixels around the intensity maxima of the probe and conjugate. We subtract the cropped regions of the two consecutive frames for the probe and conjugate to obtain the spatial intensity fluctuation images of both beams. Then, we select a region of 80$\times$80~pixels around the center of the conjugate fluctuation image and scan it over the probe fluctuation image to evaluate the spatial cross-correlation. In performing this analysis, for the far field regime we rotate the conjugate fluctuation image by 180 degrees before calculating the cross-correlation. This is due to the momentum anti-correlations that result from the phase matching condition, which in turn make the correlated regions between the probe and conjugate be diametrically opposite to each other in the far field.  The calculated spatial cross-correlations in the near and far field regimes are shown in Fig.~\ref{Fig2}.  The presence of a peak  shows the correlated region, i.e. the coherence area~\cite{Kumar3}, between the probe and conjugate spatial intensity fluctuations.  As a check, we also performed the cross-correlation measurements with images of coherent state pulses both in the near and far field and, as expected, a correlation peak was not present.

The widths of the near and far field cross-correlation peaks shown in Fig.~\ref{Fig2} give a measure of the relative uncertainty in position ($\Delta r$) and momentum ($\Delta p$), respectively, between the twin beams. To obtain a measure of these widths, we fit the cross-correlations with a two-dimensional Gaussian function of the form $Ae^{-[(x-x_0)^2/2\sigma_x^2+(y-y_0)^2/2\sigma_y^2]}$, where $A$ is a constant and $\sigma_x$  ($\sigma_y$) is the standard deviation along the $x$ ($y$) direction.
From the measured cross-correlations we obtain values of $\sigma_x$ and $\sigma_y$ in the near field of $4.27\pm0.10$~pixels and $3.52\pm0.08$~pixels, respectively, and of $4.78\pm0.13$~pixels and $4.90\pm0.13$~pixels in the far field, respectively, where the uncertainties represent the 95\% confidence intervals of the fits. To translate these results to standard deviations in actual position for the near field measurements, we take into account the demagnification factor $M=0.65$ of the imaging system, such that $\Delta r_{i}=\sigma_{i}s/M$, where $i=\{x,y\}$ and $s=16~\mu$m is the linear pixel size. Similarly, to translate to standard deviation in actual momentum for the far field measurements, we take into account the transformation $\Delta p_{i}=\frac{2\pi\hbar}{\lambda f}\sigma_{i}s$ performed by the $f$-to-$f$ optical system, where $i=\{x,y\}$,  $\lambda$ is the wavelength of the light (795~nm), and $f$ is the focal length of the Fourier lens (500~mm).

After performing the necessary transformations, we obtained the uncertainty products defined in Eq.~(\ref{Heisenberg}) along the $x$ and $y$ directions
\begin{eqnarray}
\Delta^{2} r_{x}\Delta^{2} p_{x}&=&(1.62\pm0.12)\times10^{-2}{\hbar}^2,\\
\Delta^{2} r_{y}\Delta^{2} p_{y}&=&(1.15\pm0.08)\times10^{-2}{\hbar}^2.
\end{eqnarray}
These results represent a violation of the EPR criterion by more than one order of magnitude, thus verifying the EPR paradox with quantum states of light containing a macroscopic number of photons. Furthermore, it is important to note that a statistically significant violation can be obtain with $<10$ images and in principle even with a single image (see supplemental material), as opposed to the $\sim10^{4} - 10^{5}$ images~\cite{Padgett, Lantz2, Ndagano} or long integration times~\cite{Lantz3} required for photon pair experiments.

An alternative way to verify the quantum nature of the spatial correlations is through the inseparability criterion, which is based on the total noise properties of two non-commuting observables~\cite{Duan}. For position and momentum, this criterion states that the system is entangled if the inseparability parameter $I$ satisfies the relation
\begin{equation}
   I=\langle\Delta^{2} \hat R\rangle+\langle\Delta^{2} \hat P\rangle < 2,
\end{equation}
where $\Delta^{2}\hat{R}=\Delta^{2}(\hat r_p-\hat r_c)$ and $\Delta^{2}\hat{P}=\Delta^{2}(\hat p_p + \hat p_c)$ are position difference and momentum sum variances, respectively, normalized to their corresponding shot noise.  Thus, the presence of spatial squeezing between the probe and conjugate in both the near and far fields indicates that there is spatial entanglement between them.

To show the sub-shot noise behavior, we start with the 120$\times$120~pixel cropped regions of the probe and conjugate images used for the cross-correlation measurements and align them with an image registration algorithm~\cite{Kumar2}. After the alignment, we crop a region of 80$\times$80~pixels of each probe and conjugate image around its center for the final noise analysis. We characterize the spatial quantum noise reduction (${\rm NR}$) with the ratio
\begin{equation}\label{NRF}
    {\rm NR} \equiv \frac{\langle \Delta^2[(N_{p1}-N_{p2})-(N_{c1}-N_{c2})] \rangle}{\langle N_{p1}+N_{c1}+N_{p2}+N_{c2} \rangle},
\end{equation}
where  ($N_{p1}$, $N_{c1}$) and ($N_{p2}$, $N_{c2}$) are the matrices representing the photo-counts per pixel for the cropped regions in the probe and conjugate images for the two consecutive frames used for the analysis, respectively. The statistics are calculated over the pixels of the EMCCD for each individual image consisting of two frames. Thus, the numerator represents the relative spatial variance between the probe and conjugate spatial intensity fluctuations, while the denominator represents the shot noise limit (SNL). Therefore ${\rm NR}=1$ corresponds to a coherent state and  ${\rm NR}<1$ to a spatially squeezed state. When this analysis is performed in the near (far) field configuration ${\rm NR}$ corresponds to $\langle\Delta^2 \hat R\rangle$ ($\langle\Delta^2 \hat P\rangle$).

\begin{figure}
\centering
\includegraphics{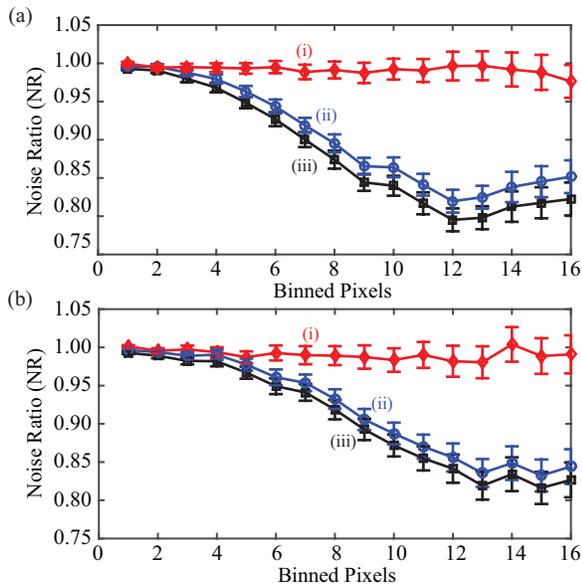}
\caption{Measured noise ratio as a function of binning in the (a) near and (b) far fields. The number of binned pixels given along the x axes represent the number of pixels used along each side of a square binning region. Traces (i), (ii), and (iii) represent the SNL, squeezing without background noise subtraction, and squeezing with background subtraction, respectively. Error bars represent the standard deviation of the mean for the noise ratio over the 200 acquired images.}
\label{Fig3}
\end{figure}
Given that the correlated regions are spread over more than one pixel, as shown in Fig.~\ref{Fig2}, an accurate measure of ${\rm NR}$ is obtained only when the noise analysis is performed after grouping or binning a square pixel region into ``super-pixels'' larger than the coherence area~\cite{Kumar1, Kumar2, Marcelo}. Figures~\ref{Fig3}(a) and (b) show NR as a function of binning in the near and far fields, respectively. For both figures, trace (i) shows the SNL, while traces (ii) and (iii) represent the noise ratios for the twin beams without and with background correction, respectively. The SNL is measured by splitting the probe laser  into two beams of equal power and performing the same noise analysis as with the bright twin beams. During these measurements the pump is turned off so that there is no FWM process or  scattered pump background noise. As expected, ${\rm NR}=1$ for the measured SNL in both the near and far field configurations. The minimum noise ratios in the near and far field configurations are 0.84$\pm$0.02 and 0.83$\pm$0.02, respectively, without background noise subtraction and 0.82$\pm$0.02 and 0.81$\pm$0.02, respectively, with background subtraction. This translates to an inseparability parameter, $I$, of $1.67\pm0.03$ without background noise subtraction and $1.63\pm0.03$ with background noise subtraction, which shows that the generated bright twin beams contain spatial entanglement.

Even though the criterion to demonstrate EPR entanglement is more stringent than the one for inseparability, we see a significantly larger violation of the EPR criterion than a reduction of the inseparability parameter. The reason for this is that the measurements performed to verify the EPR paradox are purely spatial in nature and as such directly quantify the spatial quantum properties of the system.  On the other hand, as shown in the supplemental material, the measurements to show the inseparability criterion result from an interplay between quantum correlations in the spatial and  temporal domains. More specifically, the measured level of spatial squeezing is limited by the degree of amplitude quadrature squeezing present in the twin beams. While this means that the inseparability criterion does not provide a pure measure of the degree of spatial entanglement, it does point to the presence of quantum correlations in multiple degrees of freedom (spatial and temporal) in the twin beams. This result, combined with previous measurements with the FWM source that show  the presence of quadrature entanglement in this system~\cite{Marino,Marino2}, provides a good indication of the presence of hyperentanglement in the FWM generated bright twin beams.

In conclusion, we have demonstrated the EPR paradox in its original position-momentum form  with macroscopic quantum states of light. The use of bright twin beams has made it possible for us to show a statistically significant violation of the EPR criteria with $<10$ images. We have further verified the presence of spatial entanglement through the inseparability criterion by measuring spatial squeezing in both the near and far fields. As we show, the presence of spatial squeezing results from an interplay between quantum correlations in the spatial and temporal degrees of freedom. Thus, the results presented point to the presence of hyperentanglement in the generated bright twin beams. The ability to generate  spatially entangled bright twin beams makes the FWM system a unique choice to enable quantum enhanced sensing and quantum imaging configurations that can surpass the classical state-of-the-art. The results presented thus provide a path for many novel quantum technologies to move out of the laboratory and into real-life applications.

This work was supported by the W.~M. Keck Foundation and by the National  Science  Foundation (NSF) (PHYS-1752938).

\end{document}


\title{Supplemental Material: Einstein-Podolsky-Rosen Paradox\\
with Position-Momentum Entangled Macroscopic Twin Beams}

\author{Ashok Kumar$^{1,2}$, Gaurav Nirala$^{1,3}$, and  Alberto M. Marino$^{1,3}$}
\affiliation{$^1$Homer L. Dodge Department of Physics and Astronomy, The University of Oklahoma, Norman, Oklahoma 73019, USA}
\affiliation{$^2$ Department of Physics, Indian Institute of Space Science and Technology, Thiruvananthapuram, Kerala, 695547, India}
\affiliation{$^3$Center for Quantum Research and Technology, The University of Oklahoma, Norman, Oklahoma 73019, USA}

\maketitle

\section{Statistically Significant Violation of EPR Criteria}

To determine the number of images needed to obtain a statistically significant violation of the EPR criteria, we define the confidence level parameter ($C$) along the lines of Ref.~\cite{Ndagano}
\begin{equation}
  C_{i}=\left|\frac{1/4-\Delta ^{2}r_{i}\Delta^{2} p_{i}}{\delta}\right|,
\end{equation}
where $i$ indicates the $x$ or $y$ direction and $\delta$ is the standard deviation in the estimation of the product $\Delta^{2} r_{i}\Delta^{2} p_{i}$. We consider the violation to be statistically significant when $C>5$, which represents a violation by more than 5 standard deviations.

We repeat the analysis outlined in the main text to calculate the spatial cross-correlations in the near and far fields for different number $N$ of images. We then obtain the position and momentum uncertainties, $\Delta r$ and $\Delta p$, respectively,  by fitting the calculated spatial cross-correlations with a Gaussian function of the form
\begin{equation}\label{Gauss}
  Ae^{-[(x-x_0)^2/2\sigma_x^2+(y-y_0)^2/2\sigma_y^2]},
\end{equation}
where $A$ is the amplitude and $\sigma_x$ and $\sigma_y$ are the standard deviations along the $x$ and $y$ directions, respectively, in units of pixels.  From here, we use the transformations outlined in the main text to calculate the product $\Delta^{2} r_{i}\Delta^{2} p_{i}$. To obtain $\delta$ we use the 68\% confidence intervals from the fits together with error propagation to obtain the standard deviation in the estimation of the product $\Delta^{2} r_{i}\Delta^{2} p_{i}$.  We repeat this procedure as many times as possible given the number of images $N$ used for the analysis and the 200 total images.  That is, for $N=5$ we obtain 40 different values of $C$, for $N=10$ we obtain 20 different values, and so on. Finally, for each value of $N$ we average over all the values of $C$.

\begin{figure}[h]
\centering\includegraphics{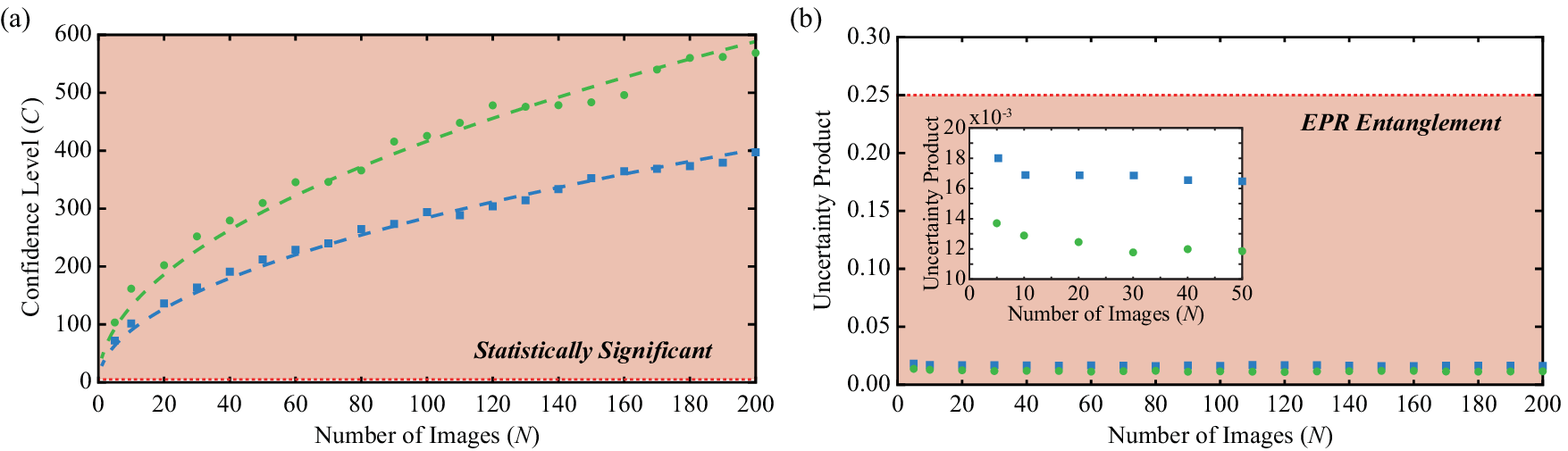}
\caption{(a) Confidence level $C$ as a function of the number of images $N$ used for the analysis. The blue squares and green circles represent the calculated values of $C$ along the $x$ and $y$ directions, respectively, while the dashed lines correspond to a fit of the form $A_{0}/\sqrt{N}$ with fit parameter $A_{0}$.  The region above the dotted red line (shaded in red) corresponds to $C>5$, which represents a statistically significant violation of the EPR criteria.
(b) Uncertainty product $\Delta^{2} r_{i}\Delta^{2} p_{i}$ as a function of the number of images $N$ used for the analysis. The region below the dotted red line (shaded in red) indicates a violation of the EPR criteria and thus the presence of EPR entanglement. The inset shows a zoom in for the region $N<50$. Measurements along the $x$ and $y$ directions are shown with blue squares and green circles, respectively.}
\label{FigS1}
\end{figure}

Figure~\ref{FigS1}(a) shows the calculated confidence level as a function of the number of images used for the analysis with  the blue squares and green circles representing the calculated values of $C$ along the $x$ and $y$ directions, respectively.  The dashed lines give a fit of the calculated values of $C$ to the function $A_{0}/\sqrt{N}$, where $A_{0}$ is a fit parameter. As expected, $C$ increases as $N$ increases and scales as $1/\sqrt{N}$. As can be seen we get a statistically significant violation of the EPR criteria even when we perform the analysis with 5 images. However, for this small number of images the fit of the cross-correlation functions to the Gaussian function of the form given in Eq.~(\ref{Gauss}) is not a good fit due to the low signal-to-noise ratio.  This results in an increase in the uncertainty product shown in Fig.~\ref{FigS1}(b) for $N=5$.  Thus, we only claim a statistically significant violation for $N<10$ in the main text. It is important to note, however, that from the scaling of $C$ with $N$ a statistically significant violation is possible even with a single image in principle, as can be seen from the fits in Fig.~\ref{FigS1}(a).

\section{Relation Between Spatial Squeezing and Temporal Squeezing}

As outlined in the main text, we characterize the spatial squeezing through the noise reduction factor (${\rm NR}$) defined as the ratio of the spatial variance of the difference between the fluctuations of the probe and conjugate to the corresponding shot noise.  We can then write ${\rm NR}$ in terms of number operators as
\begin{equation}\label{NR}
    {\rm NR} \equiv \frac{\langle \Delta^2[(\hat{N}_{p1}-\hat{N}_{p2})-(\hat{N}_{c1}-\hat{N}_{c2})] \rangle}{\langle \Delta^2[(\hat{N}_{p1}-\hat{N}_{p2})-(\hat{N}_{c1}-\hat{N}_{c2})] \rangle_{\rm CS}},
\end{equation}
where $\hat{N}_{pi}$ and $\hat{N}_{ci}$ are the number operators for the probe and conjugate images, respectively, acquired in frame $i=\{1,2\}$. In this equation the subscript ${\rm CS}$ indicates that the variance is to be evaluated for coherent states equivalent to the probe and conjugate beams. Note that for a coherent state the variance is equal to its mean, such that Eq.~(\ref{NR}) takes the form given in Eq.~(5) of the main text.  To relate this expression to the degree of temporal squeezing, we first consider the operator for the number of photons measured by the EMCCD for a given image, that is
\begin{equation}
    \hat{N}=\int_{A}d\vec{x}\int_{t_{d}}dt\:\hat{n}(\vec{x},t),
\end{equation}
where the spatial integral is over the analysis region $A$ of the images captured by the EMCCD, the temporal integral is over the detection time $t_{d}$, and $\hat{n}(\vec{x},t)$ is the spatially dependent photon flux of the beam incident on the EMMCD. In order to study the spatial properties of the light, we need to take into account the fact that the EMMCD is composed of pixels, which allows us to write
\begin{equation}
    \hat{N}=\sum_{i,j}\int_{D_{ij}}d\vec{x}\int_{t_{d}}dt\:\hat{n}(\vec{x},t)
        =\sum_{i,j}\int_{D_{ij}}d\vec{x}\int_{t_{d}}dt\:[\langle\hat{n}(\vec{x},t)\rangle +\delta\hat{n}(\vec{x},t)],
\end{equation}
where $D_{ij}$ is the area of pixel $(i,j)$ and the summation is over all the EMCCD pixels in the analysis region $A$. For the last equality we have expressed the spatially dependent flux $\hat{n}(\vec{x},t)$ as a sum of the contributions from its mean value $\langle\hat{n}(\vec{x},t)\rangle$ and its fluctuations $\delta\hat{n}(\vec{x},t)$ with $\langle\delta\hat{n}(\vec{x},t)\rangle=0$.

Next, we consider the subtraction of two subsequent frames taken a time $\Delta t$ apart
\begin{equation}\label{Ndiff}
    \delta \hat{N}=\hat{N}_{1}-\hat{N}_{2}
    =\sum_{i,j}\int_{D_{ij}}d\vec{x}\int_{t_{d}}dt\:[\langle\hat{n}_{1}(\vec{x},t)\rangle -\langle\hat{n}_{2}(\vec{x},t+\Delta t)\rangle +\delta\hat{n}_{1}(\vec{x},t)-\delta\hat{n}_{2}(\vec{x},t+\Delta t)].
\end{equation}
Here, we consider the case in which $\Delta t$ is short enough such that the spatially dependent mean value does not change from one frame to the other, but significantly longer than the inverse of the bandwidth of the process such that the quantum properties are uncorrelated both in space and time. In this case we have that
\begin{eqnarray}
    \langle \hat{n}_{1}(\vec{x},t)\rangle=\langle \hat{n}_{2}(\vec{x},t+\Delta t)\rangle,\\
    \langle \delta\hat{n}_{1}(\vec{x},t)\delta\hat{n}_{2}(\vec{x}\,',t+\Delta t)\rangle=0,\label{no_corr}
\end{eqnarray}
where we are assuming the ideal case of no technical noise. Thus, Eq.~(\ref{Ndiff}) simplifies to
\begin{eqnarray}\label{Ndiff_pix}
    \delta \hat{N}= \sum_{i,j}\int_{D_{ij}}d\vec{x}\int_{t_{d}}dt\: [\delta\hat{n}_{1}(\vec{x},t)-\delta\hat{n}_{2}(\vec{x},t+\Delta t)]
    = \sum_{i,j}\int_{t_{d}}dt\: [\delta\hat{n}^{ij}_{1}(t)-\delta\hat{n}^{ij}_{2}(t+\Delta t)],
\end{eqnarray}
where we have introduced the operator  $\delta\hat{n}^{ij}(t)=\int_{D_{ij}}d\vec{x}\:\delta\hat{n}(\vec{x},t)$, which represents the fluctuations in the number flux for pixel $(i,j)$.  Note that $\langle\delta\hat{N}\rangle=0$ and that the statistics of $\delta\hat{n}^{ij}(t)$ over all pixels in  analysis region $A$ gives a measure of the spatial properties of the beam.

We first look at the numerator in Eq.~(\ref{NR}), which can be rewritten as
\begin{equation}\label{Num}
    \langle \Delta^2[(\hat{N}_{p1}-\hat{N}_{p2})-(\hat{N}_{c1}-\hat{N}_{c2})] \rangle=\langle \Delta^2(\delta\hat{N}_{p}-\delta\hat{N}_{c}) \rangle=\langle (\delta\hat{N}_{p})^{2} \rangle+\langle (\delta\hat{N}_{c})^{2} \rangle-2\langle \delta\hat{N}_{p}\delta\hat{N}_{c} \rangle.
\end{equation}
From Eq.~(\ref{Ndiff}) and taking into account Eq.~(\ref{no_corr}) we can show that the first term on the right hand side, $\langle (\delta\hat{N}_{p})^{2} \rangle$, takes the form
\begin{equation}\label{noise_p}
    \langle(\delta \hat{N}_{p})^2\rangle = \sum_{i,j}\sum_{k,l}\int_{t_{d}}dt\int_{t_{d}}dt'\: [\langle\delta\hat{n}^{ij}_{p1}(t)\delta\hat{n}^{kl}_{p1}(t')\rangle
    +\langle\delta\hat{n}^{ij}_{p2}(t+\Delta t)\delta\hat{n}^{kl}_{p2}(t'+\Delta t)\rangle].
\end{equation}
In the limit in which the effective pixel area after binning the EMCCD pixels into ``super-pixels'' is larger than the coherence area, the resulting super-pixels are not correlated, which means that
\begin{equation}\label{pspec}
    \langle\delta\hat{n}^{ij}_{p}(t)\delta\hat{n}^{kl}_{p}(t')\rangle=\delta_{i,k}\delta_{j,l}\langle\delta\hat{n}^{ij}_{p}(t)\delta\hat{n}^{ij}_{p}(t')\rangle
    =\frac{\delta_{i,k}\delta_{j,l}}{2\pi}\int_{-\infty}^{\infty} d\Omega e^{-i\Omega\tau}S^{ij}_{p}(\Omega),
\end{equation}
where we have used the fact that for a stationary process the two time correlation function is equal to the Fourier transform of the power spectrum, $S^{ij}_{p}(\Omega)$, with $\tau=t'-t$. Thus, through the use of Eq.~(\ref{pspec}) and the fact that for a stationary process the correlation function only depends on the time difference, Eq.~(\ref{noise_p}) takes the form
\begin{equation}\label{varnp}
    \langle(\delta \hat{N}_{p})^2\rangle =\frac{1}{\pi} \sum_{i,j}\int_{-\infty}^{\infty} d\Omega S^{ij}_{p}(\Omega) \int_{t_{d}}dt\int_{t_{d}}dt'\: e^{-i\Omega(t'-t)}.
\end{equation}
Finally, in our experiments the effective integration time is determined by the temporal profile of the input probe pulse. As a result, for an intensity temporal profile $f(t)$ of the input probe, we can write Eq.~(\ref{varnp}) as
\begin{equation}
    \langle(\delta \hat{N}_{p})^2\rangle = \frac{1}{\pi}\sum_{i,j}\int_{-\infty}^{\infty}d\Omega S^{ij}_{p}(\Omega) \int_{-\infty}^{\infty}dt\:f(t)e^{i\Omega t}\int_{-\infty}^{\infty}dt'\:f(t')e^{-i\Omega t'}=\frac{1}{\pi}\sum_{i,j}\int_{-\infty}^{\infty}d\Omega |F(\Omega)|^{2}S^{ij}_{p}(\Omega),
\end{equation}
where $F(\Omega)$ is the Fourier transform of $f(t)$.  As can be seen from this result the spatial variance for the probe is given by the quadrature sum of the noise (power spectrum) over all pixels integrated over a frequency range determined by the magnitude squared of the Fourier transform of the intensity temporal profile of the input seed probe pulse.  This is to be expected, as the super-pixels are uncorrelated in the limit of a binning area larger than the coherence area. Following a similar procedure for the other two terms of  Eq.~(\ref{Num}), we find that numerator of Eq.~(\ref{NR}) takes the form
\begin{equation}\label{numerator}
    \langle \Delta^2[(N_{p1}-N_{p2})-(N_{c1}-N_{c2})] \rangle = \frac{1}{\pi} \sum_{i,j}\int_{-\infty}^{\infty}d\Omega |F(\Omega)|^{2}[S^{ij}_{p}(\Omega)+S^{ij}_{c}(\Omega)-2S^{ij}_{p,c}(\Omega)],
\end{equation}
where $S^{ij}_{c}(\Omega)$ is the power spectrum for the conjugate and $S^{ij}_{p,c}(\Omega)$ is the cross probe-conjugate power spectrum.

For the denominator, we can use as a starting point Eq.~(\ref{numerator}) and specialize to the case in which the probe and conjugate beams are replaced with coherent states of equal power and spatial profile.  In this case the cross power spectrum vanishes as the two coherent states are uncorrelated and the denominator thus takes the form
\begin{equation}
    \langle \Delta^2[(N_{p1}-N_{p2})-(N_{c1}-N_{c2})] \rangle_{\rm CS}= \frac{1}{\pi}\sum_{i,j}(S^{ij}_{{\rm SN},p}+S^{ij}_{{\rm SN},c})\int_{-\infty}^{\infty}d\Omega |F(\Omega)|^{2},
\end{equation}
where we have used the fact that the shot noise is white noise (i.e. independent of frequency) and $S^{ij}_{{\rm SN},p}$ ($S^{ij}_{{\rm SN},c}$) is the shot noise level for the probe (conjugate) for pixel $(i,j)$.  Finally, taking into account that for a coherent state there are no correlations between the spatial and temporal degrees of freedom, such that the shot noise power spectrum is the same for all pixels, we find that the denominator takes the form
\begin{equation}\label{denominator}
    \langle \Delta^2[(N_{p1}-N_{p2})-(N_{c1}-N_{c2})] \rangle_{\rm CS}=\frac{M_{x}M_{y}}{\pi} (S_{{\rm SN},p}+S_{{\rm SN},c})\int_{-\infty}^{\infty}d\Omega |F(\Omega)|^{2},
\end{equation}
where $M_{x}$ ($M_{y}$) is the number of pixels in the analysis region along the $x$ ($y$) direction.

From Eqs.~(\ref{numerator}) and~(\ref{denominator}) we have that the noise ratio takes the form
\begin{equation}
    {\rm NR} = \frac{1}{M_{x}M_{y}}\sum_{i,j}\int_{-\infty}^{\infty}d\Omega \left( \frac{|F(\Omega)|^{2}}{\int_{-\infty}^{\infty}d\Omega' |F(\Omega')|^{2}}\right)\left[\frac{S^{ij}_{p}(\Omega)+S^{ij}_{c}(\Omega)-2S^{ij}_{p,c}(\Omega)}{S_{{\rm SN},p}+S_{{\rm SN},c}}\right]
    = \frac{1}{M_{x}M_{y}}\sum_{i,j}\int_{-\infty}^{\infty}d\Omega\:  G(\Omega)S_{\rm diff}^{ij}(\Omega),
\end{equation}
where $S^{ij}_{\rm diff}(\Omega) \equiv[S^{ij}_{p}(\Omega) +S^{ij}_{c}(\Omega) -2S^{ij}_{p,c}(\Omega)]/[S_{{\rm SN},p}+S_{{\rm SN},c}]$ is the normalized intensity difference noise power spectrum and $ G(\Omega)\equiv|F(\Omega)|^{2}/\int_{-\infty}^{\infty}d\Omega |F(\Omega)|^{2}$ effectively acts as a normalized frequency filter that selects the portion of the squeezing spectrum to integrate over.  As can be seen from this expression, spatial squeezing results from the presence of temporal squeezing in the twin beams.  In fact, in the limit that the area covered by the binned pixels (super-pixel) is larger than the coherence area, the noise reduction factor ${\rm NR}$ will be equal to the average over all pixels of the normalized intensity difference noise spectra integrated over the frequency region determined by $G(\Omega)$.

\begin{figure}[h]
\centering\includegraphics{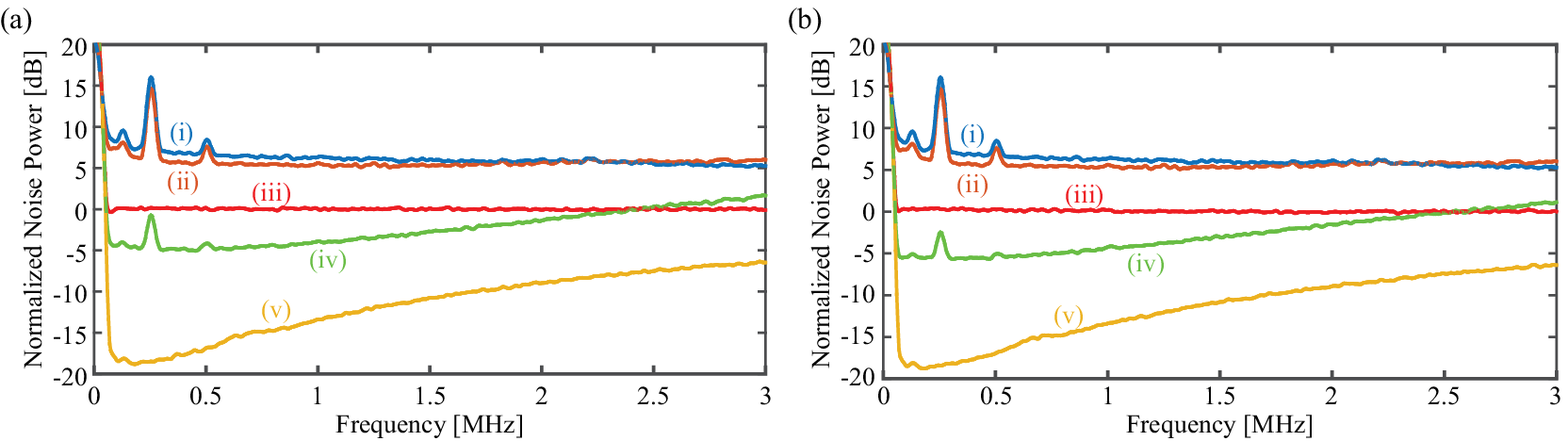}
\caption{Normalized noise power spectra for the (a) near field and (b) far field configurations. The different traces represent the normalized noise spectra for the (i) probe, (ii) conjugate, and (iv) intensity difference. Trace (iii) gives the measured shot noise, which as expected is at 0~dB, and trace (v) represents the electronic noise of our detection system. }
\label{FigS2}
\end{figure}

For our experiment the normalized intensity difference  noise power spectra for the near and far field configurations are shown in Figs.~\ref{FigS2}(a) and~\ref{FigS2}(b), respectively. For both configurations we have measured the probe (trace i) and  conjugate (trace ii) noise spectra, shot noise (trace iii), probe and conjugate intensity difference noise spectrum (trace iv), and electronic noise (trace v). All noise traces have been normalized to their corresponding shot noise. As can be seen from these figures, we have a maximum intensity difference squeezing of 5.07~dB and 5.75~dB in the near and far field configurations, respectively. While in principle we should expect the spatial squeezing to saturate at levels close to the maximum intensity difference squeezing as a function of binning, our measured degree of spatial squeezing saturates at 1~dB.  This is most likely due to a combination of the reduced quantum efficiency of the EMCCD ($\sim 70\%$) with respect to the photodiodes used to measure the intensity difference squeezing ($\sim 95\%$) and the fact that the two frame subtraction procedure that we implement is not able to cancel out all the low frequency classical technical noise present in the twin beams.